\documentclass[12pt]{amsart}

\usepackage{amsfonts}
\usepackage{amsmath,amscd,lscape}
\usepackage{amsthm}
\usepackage{amssymb}
\usepackage{latexsym}
\usepackage{mathrsfs}
\usepackage{tikz}

\usepackage{color,ulem}

\newcommand{\bea}{\begin{eqnarray}}
\newcommand{\eea}{\end{eqnarray}}
\newcommand{\beano}{\begin{eqnarray*}}
\newcommand{\eeano}{\end{eqnarray*}}
\newcommand{\beq}{\begin{equation}}
\newcommand{\eeq}{\end{equation}}

\newcommand{\mb}[1]{\hspace{2.1ex}\mbox{#1}\hspace{2.1ex}}
\numberwithin{equation}{section}

    \def\cB{{\mathcal B}}

    \def\cW{{\mathcal W}}

\newcommand{\CC}{{\mathbb C}}

\newcommand{\un}{\mathbb{I}}

\numberwithin{equation}{section}
\begin{document}

\title[Reflection and Hahn algebras  ]{Truncation of the reflection algebra\\[2mm] and the Hahn algebra}

\author{Nicolas Cramp\'e$^{\dagger,*}$}
\address{$^\dagger$ Institut Denis-Poisson CNRS/UMR 7013 - Universit\'e de Tours - Universit\'e d’Orl\'eans; 
Parc de Grammont, 37200 Tours, France.}
\email{crampe1977@gmail.com}

\author{Eric Ragoucy$^{\ddagger}$}
\address{$^\ddagger$ Laboratoire de Physique Th{\'e}orique LAPTh;
  CNRS and Universit{\'e} de Savoie;
  9 chemin de Bellevue, BP 110, F-74941  Annecy-le-Vieux Cedex, France.}
\email{eric.ragoucy@lapth.cnrs.fr}

\author{Luc Vinet$^{*}$}
\address{$^*$ Centre de recherches, math\'ematiques, Universit\'e de Montr\'eal; 
P.O. Box 6128, Centre-ville Station;
Montr\'eal (Qu\'ebec), H3C 3J7, Canada.}
\email{vinet@CRM.UMontreal.ca}

\author{Alexei Zhedanov$^{**}$}
\address{$^{**}$ Department of Mathematics, School of Information;
Renmin University of China;
Beijing 100872, China.}
\email{zhedanov@yahoo.com}

\begin{abstract}
In the context of connections between algebras coming from quantum integrable systems and algebras associated to the
orthogonal polynomials of the Askey scheme, we prove that the truncated reflection algebra attached to the Yangian 
of $sl_2$ is isomorphic to the Hahn algebra.
As a by-product, we provide a general set-up based on Euler polynomials to study truncations of reflection algebras.
\end{abstract}

\maketitle

%

%
\vspace{0cm}

\vspace{3mm}

\section{Introduction}

The purpose of this letter is to establish a connection between the algebras associated to the Askey scheme and some subalgebras of quantum groups used in 
the study of quantum integrable systems.   
On the one hand, it is by now well established that the bispectral properties of the polynomials of the Askey scheme \cite{KLS} can be encoded in 
algebras of quadratic type. These algebras are referred to as the Racah and Askey-Wilson algebras \cite{Z91,GZ, GLZ} and bear
the names of the associated polynomials at the top of the hierarchies. The algebras associated to the other families are similarly identified by the names of the corresponding polynomials. 
On the other hand, in the context of the quantum inverse scattering method \cite{STF} developed to compute the spectrum of integrable systems,
deformations of the Lie algebras called quantum groups \cite{Dr} have been studied intensively. In particular, the so-called reflection algebras which are
coideal of quantum groups, play an important role in the study of integrable systems with boundaries.

Surprisingly, these two types of algebras are connected. The Askey-Wilson algebra is a subalgebra of the reflection algebra 
associated to the quantum group $U_q(\widehat{sl}_2)$ \cite{Bas05,BK05a,BK05b}. Similar results have also been established 
for the non deformed $U(sl_2[z])$ \cite{BBC17,BC18}.
In this letter, we explore this connection further and show that a subalgebra of the reflection algebra associated to the Yangian of $sl_2$ is the Hahn algebra which is related 
to the eponym polynomials of the Askey scheme. 

The plan of the letter is as follows. In Section \ref{sec:RA}, we recall the definition of the reflection algebra $\cB(2,1)$  
associated to the Yangian of $sl_2$ and
give different properties used subsequently. In Section \ref{sec:tr}, for a given integer $N$, we define subalgebras of $\cB(2,1)$ 
called truncated reflection algebra of level $N$ and 
provide a realization of this truncation in terms of elements of $U(sl_2)^{\otimes N}$. In Section \ref{sec:tr1}, we study the 
truncation at the level 1 and
prove an isomorphism with the Hahn algebra. In Section \ref{sec:tr2}, the truncation at the level 2 is described.
We conclude by pointing out interesting links with $\cW$ algebras and with perspectives.

\section{Reflection algebra $\cB(2,1)$ \label{sec:RA}}

In this section, we recall briefly the definition and some properties of the reflection algebra $\cB(2,1)$  (see e.g. \cite{MR} for more details).

\subsection{FRT presentation.}
The key element of the FRT presentation is the $R$-matrix. In this letter, we focus on the well-known 6-vertex rational R-matrix given by
\begin{equation}
 R(x)=\begin{pmatrix}
      1-\frac{1}{x} & 0  & 0 &0\\
      0& 1& -\frac{1}{x} & 0\\
      0 &-\frac{1}{x}&1 &0\\
      0&0&0&  1-\frac{1}{x}
     \end{pmatrix}\;.
\end{equation}
It satisfies the Yang--Baxter equation
\begin{eqnarray}
&&R_{12}(x-y)R_{13}(x-z)R_{23}(y-z)=R_{23}(y-z)R_{13}(x-z)R_{12}(x-y)_ ,
\end{eqnarray}
and allows to define the Yangian of $sl_2$ \cite{Dr}.
For this R-matrix, the reflection equation \cite{Skl,Cher} is
\begin{equation}
R(x-y)\, B_1(x)\, R(x+y)\, B_2(y) = B_2(y)\, R(x+y)\, B_1(x)\, R(x-y)\ ,\label{RE}
\end{equation}
where $B_1(x)= B(x)\otimes \un_2$ and $B_2(x)=\un_2 \otimes B(x)$ with $\un_2$ the 2 by 2 identity matrix.
If we set 
\begin{equation}
 B(x)=\begin{pmatrix}
       1&0\\
       0& -1
      \end{pmatrix}+
      \sum_{n=1}^{\infty}\frac{1}{x^n}
   \begin{pmatrix}
       b_{11}^{(n)} & b_{12}^{(n)} \\
       b_{21}^{(n)} &b_{22}^{(n)}
      \end{pmatrix}\ ,\label{eq:B}
\end{equation}
the reflection equation \eqref{RE} provides the defining commutation relations between the generators $b_{ij}^{(n)}$ of the reflection algebra $\cB(2,1)$ \cite{MR}.
The Yang-Baxter equation satisfied by the R-matrix ensures that this algebra is associative.

\subsection{Unitarity.} It is proved in \cite{MR} that the unitarity relation
\begin{equation}
 B(x)B(-x)=\delta(x)\ \un_2\ ,\label{eq:unit}
\end{equation}
holds in $\cB(2,1)$ and it is further shown that the coefficients of the series $\delta(x)$ are central in $\cB(2,1)$.

\subsection{Dressing procedure.}
The dressing procedure consists in obtaining a new solution of the reflection equation from a known one. In this context,
the reflection algebra $\cB(2,1)$ can be realized in terms of the generators of the Lie algebra $sl_2$.
Let us introduce
\begin{equation}
 L(x)=\frac{1}{2}\begin{pmatrix}
                     2x-1-h & -2f\\
                    -2e& 2x-1+h
                    \end{pmatrix}\label{eq:L}
\end{equation}
where $\{h ,e, f\}$ are the generators of $sl_2$ satisfying
\begin{equation}
 {}[h,e]=2e \ , \quad [h,f]=-2f \ ,\quad  [e,f]=h \ .
\end{equation}
The quadratic Casimir elements of $sl_2$ is $\mathfrak{c}=2\{e,f\}+h^2$.
The matrix $L(x)$ satisfies the relations
\begin{equation}
 R(x-y)L_1(x)L_2(y)=L_2(y)L_1(x)R(x-y) \ , \quad  L_2(y)R(x+y)L_1(x)=L_1(x)R(x+y)L_2(y)\;.
\end{equation}
Let us remark that $L(x)L(-x)=(-x^2+\frac{1+\mathfrak{c}}{4})\un_2$.

For any solution $B(x)$ of the reflection equation, we can construct 
\begin{equation}
 L(x)B(x)L(x)
\end{equation}
which is also a solution of the reflection equation if $[B_1(x),L_2(y)]=0$, that is if any entries of $B(x)$ 
commute with any entry of $L(x)$. This is the case if the entry of $B(x)$ are scalar or if the entries of $B(x)$ and $L(x)$ are 
in different factors of a tensor product.

\subsection{Serre--Chevalley presentation.} In \cite{BC12}, an alternative presentation of $\cB(2,1)$ is obtained. The reflection algebra $\cB(2,1)$ is generated 
by $H$, $E$ and $F$ subject to the following relations 
\begin{eqnarray}
&& [H,E]=2E \ , \qquad [H,F]=-2F\ ,\label{SC1}\\
&& [E,[E,[E,F]]]=-12 EHE\ , \qquad [F,[F,[F,E]]]=12FHF\ .\label{SC2}
\end{eqnarray}
The relations \eqref{SC2} can be viewed as the Serre relations for the $\cB(2,1)$ algebra.

\section{Truncation $\cB^{(N)}(2,1)$ of $\cB(2,1)$ \label{sec:tr}}

In the previous section, we have presented the reflection algebra $\cB(2,1)$ and shown that its defining relations 
can be given in the FRT presentation with the  generators collected in the series \eqref{eq:B}. 
We are now interested in the case when the series are truncated and become polynomials. 
For convenience, instead of polynomials in $\frac1x$ we will consider polynomials in $x$. Indeed, for any truncation of $\cB(2,1)$, 
we can always multiply $B(x)$ by an appropriate power of $x$ to get an equivalent description of the truncated algebra using polynomials in $x$. It is this formulation we adopt from now on.

\subsection{Definition of $\cB^{(N)}(2,1)$.} Let $N\in\mathbb{Z}_{\geq 0}$. The truncated reflection algebra $\cB^{(N)}(2,1)$ are generated 
by $\mu^{(N)}$, $h^{(N)}_{2n}$, $\overline h^{(N)}_{2n}$, $e^{(N)}_{2n+1}$ and $f^{(N)}_{2n+1}$ for $n=0,1,\dots N-1$. 
The generators $\mu^{(N)}$ are central in  $\cB^{(N)}(2,1)$.
We organize the generators in the following matrix
\begin{equation}
 B^{(N)}(x)=\begin{pmatrix}
      x\ \overline h^{(N)}(x) -h^{(N)}(x) & f^{(N)}(x) \\
       e^{(N)}(x) & -x\ \overline h^{(N)}(x)-h^{(N)}(x)
      \end{pmatrix} \label{eq:BN}
\end{equation}
where we have defined the polynomials
\begin{eqnarray}
 h^{(N)}(x)&=&\sum_{n=0}^N E_{2N-2n}(x)\ h^{(N)}_{2n}
 {\quad \mb{with} h^{(N)}_{2N}=\mu^{(N)}}\ ,\label{eq:hNx}\\
 \overline h^{(N)}(x)&=&\sum_{n=0}^{N} E_{2N-2n}(x)\ \overline{h}^{(N)}_{2n-2}
{\quad \mb{with}\overline h^{(N)}_{-2}=1}\ ,\\
 f^{(N)}(x)&=&\frac{2}{x-1}\,\sum_{n=0}^{N-1} E_{2N-2n}(x)\ f^{(N)}_{2n+1}
\quad \mb{with} f^{(N)}(0)=0 \ , \label{eq:fNx}\\
 e^{(N)}(x)&=&\frac{2}{1-x}\,\sum_{n=0}^{N-1} E_{2N-2n}(x)\ e^{(N)}_{2n+1}
  \quad \mb{with} e^{(N)}(0)=0\ .\label{eq:eNx}
\end{eqnarray}
where $E_{2n}(x)$ are the Euler polynomials of degree $2n$ (see \eqref{eq:defEB} for their definition).
Let us remark that the R.H.S. of \eqref{eq:fNx} and \eqref{eq:eNx} are polynomials since the Euler polynomials are divisible by $x-1$.
The defining relations of $\cB^{(N)}(2,1)$ are given by the reflection equation \eqref{RE} applied to $B^{(N)}(x)$. 
The properties of the Euler polynomials ensure that the expansions \eqref{eq:hNx}-\eqref{eq:eNx} are consistent with \eqref{RE}, 
this will become manifest below, when we provide a realisation of these generators in $U(sl_2)^{\otimes N}$. 

For latter purpose, we need the following relation satisfied by the Euler polynomials
\begin{equation}
 x(x-1)E_{2n}(x)=E_{2n+2}(x)-2(2n)! \sum_{k=1}^n \frac{(1-2^{2k})B_{2k}}{(2n-2k+1)!(2k)!} E_{2n-2k+2}(x)\;,\label{eq:prE}
\end{equation}
where $B_n$ are the Bernoulli numbers. The proof of this relation is given in Appendix \ref{App}.

For $N=0$, the matrix \eqref{eq:BN} reduces to 
$B^{(0)}(x)=\begin{pmatrix}
                  x-\mu^{(0)}&0\\
                  0&-x-\mu^{(0)}
                 \end{pmatrix}$ and $\mu^{(0)}$ can be considered as a number.
              
The truncation $\delta^{(N)}(x)$ of the central element $\delta(x)$, \eqref{eq:unit},
\begin{equation}
 B^{(N)}(x)B^{(N)}(-x)=\delta^{(N)}(x) \un_2
\end{equation}
is central in $\cB^{(N)}(2,1)$.          
                 
\subsection{Recursive construction.} By using the dressing procedure introduced in the previous section, we can obtain a realisation of $\cB^{(N+1)}(2,1)$
in $\cB^{(N)}(2,1)\otimes U(sl_2)$. Indeed, if $B^{(N)}$ satisfy the reflection equation, then
\begin{equation}\label{eq:recur}
 B^{(N+1)}(x)=L(x)B^{(N)}(x)L(x)
\end{equation}
with $L(x)$ given by \eqref{eq:L}, satisfies also the reflection equation. More precisely, let us note 
that in \eqref{eq:recur}, the entries of $L(x)$ belong to $U(sl_2)$, while the entries $B^{(N)}(x)$ belong to $\cB^{(N)}(2,1)$.
Then, the r.h.s. of \eqref{eq:recur} belongs to $\cB^{(N)}(2,1)\otimes U(sl_2)$, and the equality provides a morphism from 
$\cB^{(N+1)}(2,1)$
to $\cB^{(N)}(2,1)\otimes U(sl_2)$.
Explicitly, one gets 
\begin{eqnarray}
 h^{(N+1)}(x)&=&x(x-1)\left( h^{(N)}(x)\otimes 1+\overline h^{(N)}(x)\otimes h\right) +(x-1)\left(f^{(N)}(x)\otimes e +e^{(N)}(x)\otimes f\right)\nonumber \\
 &&  +\frac{1}{4}\ h^{(N)}(x)\otimes (1+\mathfrak{c})\ ,\label{eq:hn1}\\
 \overline{h}^{(N+1)}(x)&=&x(x-1)\overline h^{(N)}(x)\otimes 1+\frac{1}{4}\overline h^{(N)}(x)\otimes (2h^2+4h+1-\mathfrak{c})\nonumber\\
 &&+\frac{1}{4x}\left(e^{(N)}(x)\otimes\{h,f\}+f^{(N)}(x)\otimes\{h,e\}\right) \ , \label{eq:hbn1}\\
   e^{(N+1)}(x)&=&x(x-1) e^{(N)}(x)\otimes 1 +2xh^{(N)}(x)\otimes e+\frac{x}{2}\ \overline h^{(N)}(x)\otimes \{h,e\}\nonumber \\
   &&+ f^{(N)}(x)\otimes e^2+\frac{1}{4}e^{(N)}(x)\otimes (1-h^2)\ ,\label{eq:en1}\\
    f^{(N+1)}(x)&=& x(x-1) f^{(N)}(x)\otimes 1 + 2x h^{(N)}(x)\otimes f +\frac{x}{2}\ \overline h^{(N)}(x)\otimes \{h,f\}\nonumber\\
    && +e^{(N)}(x)\otimes f^2+\frac{1}{4}f^{(N)}(x)\otimes (1-h^2)\ .\label{eq:fn1}
\end{eqnarray}
By using this procedure recursively, one can find a realisation of $\cB^{(N)}(2,1)$ in $U(sl_2)^{\otimes N}$.

Let us remark that it is not obvious that $h^{(N+1)}(x)$, $\overline h^{(N+1)}(x)$, $e^{(N+1)}(x)$ and $f^{(N+1)}(x)$ defined by the 
relations \eqref{eq:hn1}-\eqref{eq:fn1} can be expanded as in \eqref{eq:hNx}-\eqref{eq:eNx} for 
$h^{(N)}(x)$, $\overline h^{(N)}(x)$, $e^{(N)}(x)$ and $f^{(N)}(x)$ themselves given by these expansions in terms of the Euler polynomials.
However, 
by using \eqref{eq:prE}, we can show that it is the case and that $h^{(N+1)}_{2n}$, $\overline h^{(N+1)}_{2n}$, $e^{(N+1)}_{2n+1}$ 
and $f^{(N+1)}_{2n+1}$ are uniquely determined.
For example, setting $x=0$ in \eqref{eq:hn1}, we get $\mu^{(N+1)}=\frac14 \mu^{(N)}\otimes(1+\mathfrak{c})$ which is indeed central in 
$\cB^{(N)}(2,1)\otimes U(sl_2)$.

The truncation $\delta^{(N)}(x)$ also obeys a recursion: $\delta^{(N+1)}(x)=(-x^2+\frac{1+\mathfrak{c}}4)\delta^{(N)}(x)$.

\section{Truncation $\cB^{(1)}(2,1)$ of $\cB(2,1)$  \label{sec:tr1}}

In this section, we study in detail the first truncation of the reflection algebra. We set $N=1$ in $B^{(N)}(x)$.

\subsection{Definition of $\cB^{(1)}(2,1)$.}
By using the explicit form of the Euler polynomials, one gets
\begin{equation}
 B^{(1)}(x)=\begin{pmatrix}
       x^3-(h_0+1)x^2+(h_0+\overline{h}_0)x-\mu & 2xf_1 \\
       -2xe_1 & -x^3-(h_0-1)x^2+(h_0-\overline{h}_0)x-\mu
      \end{pmatrix}\,.
\end{equation}
To lighten the presentation, we have dropped the superscripts $(1)$ and used the notation $h_0$,  $\overline{h}_0$, $e_1$, $f_1$ and $\mu$ for the generators of $\cB^{(1)}(2,1)$.
The reflection equation \eqref{RE} for $B^{(1)}(x)$ is equivalent to $\mu$ being central and to the commutation relations
\begin{eqnarray}
&& [h_0,e_1]=2e_1\ , \quad [h_0,f_1]=-2f_1 \ , \quad [e_1,f_1]=-h_0\overline{h}_0+\mu\ ,\\
&& [\overline h_0,e_1]=\{h_0,e_1\}\ , \quad [\overline h_0,f_1]=-\{h_0,f_1\} \ , \quad [h_0,\overline h_0]=0\ .
\end{eqnarray}
The element $\gamma=\frac{1}{2} h_0^2-\overline h_0$ is central in $\cB^{(1)}(2,1)$. It follows that $\cB^{(1)}(2,1)$
can be generated by $h_0$, $e_1$, $f_1$ and the central element $\gamma$ subject to
\begin{eqnarray}
 && [h_0,e_1]=2e_1\ , \quad [h_0,f_1]=-2f_1 \ , \quad [e_1,f_1]=\gamma h_0-\frac{1}{2}h_0^3+\mu\ .
\end{eqnarray}
These relations are the defining relations of the Higgs algebra \cite{H79} and
we conclude therefore that $\cB^{(1)}(2,1)$ is isomorphic to this algebra. 

It is easy to show that $h_0$, $e_1$ and $f_1$ satisfy also the defining relations \eqref{SC1}-\eqref{SC2} of $\cB(2,1)$.
Indeed, one gets 
\begin{eqnarray}
 [e_1,[e_1,f_1]]=[ e_1, \gamma h_0-\frac{1}{2}h_0^3]=-2\gamma e_1 +3h_0e_1h_0+4e_1\ .
\end{eqnarray}
Then 
\begin{eqnarray}
 [e_1,[e_1,[e_1,f_1]]=3[e_1,h_0e_1h_0]=-12e_1h_0e_1
\end{eqnarray}
which is the first relation of \eqref{SC2}. The second relation is proven similarly.
This gives another proof that the Higgs algebra is a subalgebra of $\cB(2,1)$.

\subsection{Center of $\cB^{(1)}(2,1)$.} 
From the coefficient of $x^2$ in the expansion of $\delta^{(1)}(x)$, we see that
\begin{equation}
 \delta_2=2\{e_1,f_1\}-\frac{1}{4} h_0^4+(\gamma-1) h_0^2+2\mu h_0
\end{equation}
is central in $\cB^{(1)}(2,1)$. It is easy to show that $\delta_2$ and  $\gamma$ are the only independent elements in the expansion of $\delta^{(1)}(x)$. 

A PBW basis for $\cB^{(1)}(2,1)$ is then given by $\{h_0^n,\,e_1^mh_0^n,\,f_1^mh_0^n\}_{n\geq0,m\geq1}$ 
with coefficients in $\CC[\mu,\delta_2,\gamma]$. The PBW basis is graded by $h_0$, and only the elements $h_0^n$ have grade 0. Since a linear combination of these elements cannot be central,
 this shows that $\mu,\delta_2,\gamma$ generate the center of $\cB^{(1)}(2,1)$.

\subsection{Realisation in $U(sl_2)$.}
By using the recursive construction presented in Section \ref{sec:tr}, one obtains a realisation of $\cB^{(1)}(2,1)$ in terms of the 
enveloping algebra of $sl_2$. Namely, one gets
\begin{align}
&h_0=h+\mu^{(0)}\ , 
&&\overline h_0=\frac{1}{2}h^2+\mu^{(0)} h+\frac{1}{4}-\frac{1}{4}\mathfrak{c}\ ,\\
& e_1=-\mu^{(0)}e- \frac{1}{4} \{h,e \}\ , 
&&f_1=\mu^{(0)}f +\frac{1}{4}\{h,f\} \ ,\\
& \mu=\frac{\mu^{(0)}}{4}(\mathfrak{c}+1) \ , 
 &&\gamma= \frac{(\mu^{(0)})^2}{2}-\frac{1}{4}+\frac{1}{4}\mathfrak{c}\ ,\\
 &\delta_2= \frac{1}{4}\left((\mu^{(0)})^4-(\mu^{(0)})^2(3+\mathfrak{c})-\mathfrak{c}\right)\ .
\end{align}


\subsection{Hahn algebra.} 
It is proved in \cite{Z92,GVZ,FGVVZ} that the Higgs algebra is isomorphic to the Hahn algebra.
Indeed, let us define the following generators
\begin{eqnarray}
 X =\frac{1}{2}  h_0\ ,\quad Y = -\frac{1}{4}( h_0^2-2\gamma +2e_1+2f_1)=-\frac{1}{2}(\overline{h}_0+e_1+f_1)\ .
\end{eqnarray}
Then $X$ and $Y$ satisfy the following
\begin{eqnarray}
&&[[X,Y],Y]=\{X,Y\}+\frac{\mu}{2}\ ,\\
&& [X,[X,Y]]=X^2+ Y -\frac{\gamma}{2}\ ,
\end{eqnarray}
which are the canonical defining relations of the Hahn algebra.
The inverse mapping is given by
\begin{equation}
 e_1 =\frac{\gamma}{2} -X^2-Y-[X,Y]\ ,\quad f_1 =\frac{\gamma}{2} -X^2-Y+[X,Y]\quad \text{and}\quad  h_0 = 2X\ .
\end{equation}

\section{Truncation $\cB^{(2)}(2,1)$ of $\cB(2,1)$ \label{sec:tr2}}

In this section, we study in detail the truncation of the reflection algebra for $N=2$.

\subsection{Definition of $\cB^{(2)}(2,1)$.} From the explicit form of the Euler polynomials, relations \eqref{eq:hNx}-\eqref{eq:eNx} becomes for $N=2$
\begin{eqnarray}
&& h^{(2)}(x)=\mu+x(x^3-2x^2+1) h_0+x(x-1)h_2\\
&& \overline{h}^{(2)}(x)=x(x^3-2x^2+1)+x(x-1)\overline{h}_0+\overline{h}_2\\
&& e^{(2)}(x)=-2x(x^2-x-1)e_1-2xe_3\\
&& f^{(2)}(x)=2x(x^2-x-1)f_1+2xf_3
\end{eqnarray}
We have dropped out superscripts $(2)$ of the generators $h_0$, $h_2$, $\overline h_0$, $\overline h_2$,  $e_1$, $f_1$, $e_3$, $f_3$ and $\mu$.
Using $\delta^{(2)}(x)$, we can prove that 
\begin{eqnarray}
 &&\gamma_1=\overline h_0- \frac{1}{2} h_0^2 \ ,\\
 &&\gamma_2=\overline h_2-h_0h_2+h_0^2\left(1+\frac{1}{8}h_0^2+\frac{1}{2}\gamma_1\right) -\{e_1,f_1\}\ ,
\end{eqnarray}
are central in $\cB^{(2)}(2,1)$. This allows to express $\overline h_0$ and $\overline h_2$ in terms of the other generators and the central elements $\gamma_1$ and $\gamma_2$.  
Then, the reflection equation satisfied by $B^{(2)}(x)$ is equivalent to the following relations
\begin{eqnarray}
&& [h_0,e_1]=2e_1\ , \quad [h_0,f_1]=-2f_1 \ ,\\
&& [e_1,f_1]=h_2-h_0\overline{h}_0\ , \quad  [h_0,h_2]=0\ ,\\
&& [h_2,e_1]=2e_3\ , \quad [h_2,f_1]=-2f_3 \ , \quad [h_0,e_3]=2e_3\ , \quad [h_0,f_3]=-2f_3\ ,\\
&&[e_1,e_3]=0\ , \quad [f_1,f_3]=0\ , \quad [e_1,f_3]=[e_3,f_1]=h_2-h_0(\overline h_0+ \overline h_2)+\mu \, \\
&&[h_2, e_3]=2\overline{h}_0(e_3-e_1)-2\overline h_2 e_1+2e_3\ , \quad   [h_2,f_3]= 2\overline{h}_2f_1+2\overline{h}_0(f_1-f_3)-2f_3 \ ,\\
&&[e_3,f_3]=h_2-h_0\overline h_0-\overline h_2(h_0+h_2)-2f_1e_3+2f_3e_1+\mu(1+\overline h_0)\ .
\end{eqnarray}

\subsection{Central elements of $\cB^{(2)}(2,1)$.} 
 It is seen from the coefficients of $x^2$ and $x^4$ in the expansion of $\delta^{(2)}(x)$ that
\begin{eqnarray}
 &&\delta_2= 2\{e_3-e_1,f_3-f_1\} -(h_0-h_2)^2-\overline{h}_2^2+2\mu(h_2-\overline{h}_0)\ ,\\
 &&\delta_4=2\{e_1,f_3\}+2\{e_3,f_1\}-6\{e_1,f_1\} +2h_0(2h_0-2h_2+\mu) +h_2^2 +{\overline h}_0({\overline h}_0-2{\overline h}_2-2)
\end{eqnarray}
are central in $\cB^{(2)}(2,1)$.

\section{Conclusion}

We have pursued in this letter the examination of the connection between the algebras entering the study quantum integrable systems 
and the ones associated to the orthogonal polynomials of the Askey scheme. In particular, we have proved that the truncated reflection 
algebra attached to the Yangian of $sl_2$ is isomorphic to the Hahn algebra.
This brings up many interesting questions. As shown in this letter, we can naturally construct a family of truncated reflection algebras. 
However, we succeeded only in identifying the first one
as an algebra associated to orthogonal polynomials. Could the other ones be connected to the so-called higher rank Racah or Askey-Wilson 
algebras \cite{DV,PW}? 
It would also be natural to study this correspondence for the truncated reflection algebra associated to the Yangians of $sl_n$, $so_n$ or $sp_n$. 
A first attempt in this direction has been done for
the non-deformed case in \cite{BCP}.

Note also that it could be relevant to study the limit $N\to\infty$ of $\cB^{(N)}(2,1)$. Indeed, although the limit of the truncated algebra gives back 
the original $\cB(2,1)$ reflection algebra,
the fact that we have renormalized by a power of $x$ (as explained at the beginning of Section \ref{sec:tr}) could lead to a new algebraic structure for the generators 
\eqref{eq:hNx}--\eqref{eq:eNx}.

To conclude, let us finally mention that the type of algebras studied here is also linked to $\cW$ algebras. 
The classical truncated twisted Yangian $Y_p^+(2n)$, where $Y^+(2n)$ is the twisted Yangian \cite{MNO} 
associated to $Y(gl_{2n})$  based on $so_{2n}$
and where the subscript $p$ indicates a 
truncation at level $p$, is isomorphic to the finite $\cW$ algebra $\cW(so_{2np},n\,sl_p)$ \cite{R00}.  
Since for $sl_2$, the twisted Yangian and the reflection algebra are isomorphic \cite{MR}, 
the truncations of the reflection algebra $\cB^{(N)}(2,1)$ studied here may be the quantification 
of $\cW(so_{4N+2},\, sl_{2N+1})$ (and $\cW(so_{4N},\, sl_{2N})$ for $\mu^{(N)}=0$). 
We hope to return to these matters in the future. 

\medskip

 \textbf{Acknowledgments:} N.C. is gratefully holding a CRM--Simons professorship. 
 E.R. warmly thanks the Centre de Recherches Math\'ematiques (CRM) for hospitality and support during his visit to Montreal in the 
 course of this investigation. 
 The research of L.V. is supported in part by a Natural Science and Engineering Council (NSERC) of Canada
discovery grant and that of A.Z. by the National Science Foundation of China (Grant No. 11711015).

\appendix

\section{Proof of the relation \eqref{eq:prE} \label{App}}

Let us recall that the generating functions for the Euler polynomials and the Bernoulli numbers are
\begin{equation}
 f(w)=\frac{2e^{xw}}{e^w+1}=\sum_{n=0}^\infty \frac{w^n}{n!} E_n(x)\ , \quad g(w)=\frac{w}{e^w-1}=\sum_{n=0}^\infty \frac{w^n}{n!} B_n\ .\label{eq:defEB}
\end{equation}
From the following functional relation
\begin{equation}
 f''(w)+f'(w)-x(x-1)f(w)-\frac{2}{w}(g(w)-g(2w))f'(w)=0\ ,
\end{equation}
we deduce that
\begin{equation}
 E_{n+2}(x)+E_{n+1}(x)-x(x-1)E_n(x)-2 n! \sum_{k=0}^{n} \frac{(1-2^{k+1})B_{k+1}}{(n-k)!(k+1)!} \ E_{n-k+1}(x)=0
\end{equation}

We prove \eqref{eq:prE} from the property $B_{2k+1}=0$ for $k=1,2,\dots$.

\end{document}